\documentstyle[12pt]{article}
\huge
\renewcommand{\theequation}{\arabic{section}.\arabic{equation}} 
\def\setzero{\setcounter{equation}{0}}

\newcounter{eqalph}
\def\bph{\setcounter{eqalph}{\value{equation}}
   \addtocounter{eqalph}{1}
   \setcounter{equation}{0}
   \renewcommand{\theequation}{\arabic{section}.\arabic{eqalph}\alph{equation}}}
\def\eph{\setcounter{equation}{\value{eqalph}}
   \renewcommand{\theequation}{\arabic{section}.\arabic{equation}}
\par\noindent}

\begin{document}

\baselineskip 18pt
\newcommand{\ii}{\mbox{i}}
\def \sech{{\rm sech}}
\def \tanh{{\rm tanh}}
\def \cn{{\rm cn}}
\def \sn{{\rm sn}}
\def\bm#1{\mbox{\boldmath $#1$}}
\newfont{\bg}{cmr10 scaled\magstep4}
\newcommand{\bzr}{\smash{\hbox{\bg 0}}}
\newcommand{\bzl}{%
   \smash{\lower1.7ex\hbox{\bg 0}}}
\title{Chiral solitons from dimensional reduction of 
Chern-Simons gauged  coupled non-linear
 Schr\"{o}dinger model} 
\date{\today}
\author{ Masato {\sc Hisakado}  
\\
\bigskip
\\
{\small\it Department of Pure and Applied Sciences,}\\
{\small\it University of Tokyo,}\\
{\small\it 3-8-1 Komaba, Meguro-ku, Tokyo, 153, Japan}}
\maketitle

\vspace{20 mm}

Abstract


The soliton structure of a gauge theory 
 proposed  to describe  chiral excitations 
in the multi-Layer
Fractional Quantum Hall Effect 
 is  investigated.
 A new type of derivative multi-component
 nonlinear  Schr\"{o}dinger equation emerges as  effective description of the system that supports novel chiral solitons.
We discuss the classical properties of the solutions and 
study  relations to integrable systems.
\vfill
\par\noindent
{\bf  }

\newpage

\section{Introduction}
Study of two-dimensional electron system in a strong magnetic 
field is  one of the most interesting in the condensed matter 
physics.  
In the fractional quantum hall effect, early work by 
Halperin anticipated novel fractional quantum hall 
effects due to inter-layer correlations in multi-layer systems.\cite{ha}
Recent technological progress has made it possible  to produce 
double-layer two dimensional electron gas systems  of extremely high mobility 
in which these effects can be observed.\cite{ex}

One of the most fascinating aspects of classical field theory 
is the existence  of traveling  localized  solutions of the 
nonlinear equations describing physical systems.
From the quantum field theory point of view,
 they are believed to carry information about the non-perturbative 
structure of the quantum field theory.
Their particle-like properties  have an  intuitive  and 
reasonable  interpretation 
as  bound states  of the elementary excitations  
of the corresponding quantum field theory.

Solitons made their appearance  in a completely 
different context some years ago, 
in low energy phenomenological applications 
to  physical systems confined  to a plane.
An interesting class of gauge theoretical models,
describing matter coupled to a Chern-Simons gauge field 
was introduced to obtain a simple  realization of interacting anyons.
\cite{JAPI1}
The physical  reason lies that some characteristic of the model,
 in particular its fractional statistics, can be maintained by the related
  one-dimensional excitations and that, by a
 suitable modification, chiral behavior can be 
induced.
This fact is  relevant in the phenomenological 
description of the edge states in the quantum hall effect.
\cite{RABE},\cite{WEN}
As first observed in \cite{NOI} a novel and interesting soliton 
structure  is present there, finding its origin in 
the gauge 
coupling and in the chiral modification.
Our investigation is a extension of the work begun in \cite{NOI}.

We study a family of 1+1 dimensional theories 
that describes  non relativistic bosons 
interacting with a gauge potential 
in a multi-component model.
We show the chiral excitations
which can be seen  in  one component case.
This fact is  relevant  to describe 
the edge states in the multi-Layer quantum hall effect.
In the multi-component case
 the theory contains 
 self and inter-component interactions.
We show the explicit soliton solutions which describe 
the soliton-soliton scattering for 
the inter-component interactions.
In the one component case 
if the theory is modified by adding the suitable  potential,
the model becomes  an integrable 
derivative nonlinear Schr\"{o}dinger  equation.\cite{COREA}
The solitons are no longer chiral.
In the multi-component case
even if we add   potential, 
we can not make this model integrable.

This letter is organized as follows.
We start in  section 2 by describing the 
 extension of  Jackiw-Pi model.
We show that the system is equivalent 
to a general family of coupled non-linear Schr\"{o}dinger 
equation that does  not possess Galileian invariance.
In the section 3
we discuss the conserved charges and  derive  some general 
properties of the localized classical solution.
In the section 4 we compute the soliton 
solutions  and discuss their properties.
The last section is devoted to the concluding remarks.

\setzero

\section{Nonlinear Derivative Schr\"{o}dinger  Equation 
from a Gauge Theory}

A non-relativistic gauge field theory that leads to planar
 anyons in the coupled nonlinear Schr\"{o}dinger equation,
gauged by a Chern-Simons field and governed by the Lagrange
density

\begin{equation}
{\cal L}_{(2+1)} = \frac{1}{4\bar\kappa}\epsilon^{\alpha\beta\gamma} 
\hat A_\alpha \hat F_{\beta\gamma}
 + i\hbar \sum_{k}^{N}\Psi^*_{k} (\partial_t + \ii\hat A_0)\Psi_{k} - 
\frac{\hbar^2}{2m} \sum_{k=1}^{N}\sum_{i=1}^2 \left| (\partial_i + \ii\hat A_i)\Psi_{k}\right|^2
- V.
\label{eq:1} \\
\end{equation}

Here $\Psi_{k}$ is the Schr\"{o}dinger quantum field
in the $k$-th component, giving rise to charged bosonic 
particles after second quantization. $\hat A_\mu$ possesses no propagating 
degrees of freedom; it 
can be eliminated, leaving a statistical Aharonov-Bohm interaction  between the
particles. $V$ describes possible nonlinear self-interactions and  
interactions between components.
It  can be a general polynomial in the density 
$\Psi^{*}_{k}\Psi_{k}\;\;k=1,2,\cdots N$.
$N$ is the number of  components.  We notice that 
the above Lagrangian is invariant under Galilean transformations, due to the 
topological nature of the Chern-Simons action. 
When analyzing the lineal problem \cite{NOI}, it is natural to
consider a dimensional 
reduction of (\ref{eq:1}), by suppressing dependence on the second spatial 
coordinate, renaming $\hat A_2$ as $(mc/\hbar^2) B$ and redefining the gauge 
field as $\hat A_x=A_x$ and $\hat A_0=A_0+m c^2/2\hbar B^2$. 
In this way one is led to a $B$--$F$ gauge theory coupled to a
non-relativistic bosonic fields  in $1+1$ dimensions:  
\begin{equation} 
 {\cal L}_{(1+1)} = \frac1{2\kappa} B \epsilon^{\mu\nu} F_{\mu\nu}
+ \ii\hbar \sum_{k=1}^{N}\Psi^*_{k}(\partial_t + \ii A_0)\Psi_{k} - 
\frac{\hbar^2}{2m} \sum_{k=1}^{N}\left| (\partial_x + \ii A_1)\Psi_{k}\right|^2
- V,\label{eq:2}
\end{equation}
where $\kappa \equiv (\hbar^2/mc)\bar\kappa$ is 
dimensionless and we have neglected $\partial_x (B^3/3\hbar \kappa)$ 
since it is a total spatial derivative.
 In the following 
we prefer a simpler expression that describes ``chiral'' Bose fields,
propagating only in one direction
we choose the $B$-kinetic Lagrange density to be
\begin{equation}
{\cal L}_B = {\lambda\over 2\kappa^2\hbar} \dot B B',
 \label{eq:3}
\end{equation}
and the total Lagrange density is ${\cal L} = {\cal L}_B +
{\cal L}_{(1+1)}.$ \cite{NOI}  The  equations of motion are
\bph
\begin{eqnarray}
\label{eq:5a}
&&\ii\hbar(\partial_{t}+\ii A_{0})\Psi_{i}+{\hbar^2\over 2m}
(\partial_{x}+\ii A_{1})^2\Psi_{i}-V'\Psi_{i}=0,\\
&& F_{01}-\frac{\lambda}{\kappa\hbar} \dot B'=0,
\label{eq:5b}\\
&&B'-\hbar \kappa \sum_{k=1}^{N}\Psi^{*}_{k}\Psi_{k}=0,
\label{eq:5c}\\
&&\dot B+\hbar \kappa  \sum_{k=1}^{N}\hat J_{k}=0,
\label{eq:5d}
\end{eqnarray}
\eph
with $\hat J_{k}={\hbar\over 2\ii m}\Bigl(\Psi^{*}_{k}(\partial_{x} +\ii A_{1})
\Psi_{k}
-\Psi_{k}(\partial_{x}-\ii A_{1})\Psi^{*}_{k}\Bigr)$.
Here dot/prime indicate
differentiation with respect to time/space.
  The integrability
condition for the last two equation leads to the usual  continuity 
equation 
\begin{equation}
\partial_{t}(\sum_{k=1}^{N}\Psi^{*}_{k}\Psi_{k})
+\partial_{x}\sum_{k=1}^{N}\hat J_{k}=0. 
\label{eq:7}
\end{equation}
Now in terms of  field
$\displaystyle{
\hat\phi_{i}=\exp\left (\ii \int^{x}_{x^0}~dy~ A_{1}(y,t)
+\ii \int^{t}_{t^0}~dt^\prime~ A_{0}(x^0,t^\prime)
-\ii\frac{\lambda}{\kappa\hbar}B(x^0,t)\right)\Psi_{i}}$
 (\ref{eq:5a}) and (\ref{eq:5b}) become respectively
\begin{eqnarray}
&&\ii\hbar\partial_{t}\hat\phi_{i}+{\hbar^2\over 2m}\partial_{x}^2\hat\phi_{i}-\hbar 
{\lambda}(\sum_{k=1}^{N}j_{k})\hat\phi_{i}-V'\hat\phi_{i}=0,
\label{eq:8}\\
&&F_{01}={\lambda}\partial_{t}(\sum_{k=1}^{N}\hat\phi^{*}_{k}\hat\phi_{k}).
\label{eq:9}
\end{eqnarray}
The latter gives the electro-magnetic field as a function of 
the density $\sum_{k}^{N}\hat\phi^{*}_{k}\hat\phi_{k}$ while the former encodes all the dynamical
contents  of the system.
In the following we shall be mainly interested 
in the case in which the potential 
$V$ is absent: this
 corresponds 
to a coupled Schr\"odinger equation with a {\it current\/} 
 ($\sum_{k=1}^{N}j_{k}$) nonlinearity,

\begin{equation}
j_{k}= {\hbar \over 2\ii m} \left( \hat\phi^{*}_{k}\partial_x\hat\phi_{k} -
\hat\phi_{k}\partial_x\hat\phi^{*}_{k} \right). 
\nonumber
\end{equation}

Note that 
(\ref{eq:8})  does not possess a local Lagrangian 
formulation directly in terms of the field $\hat\phi_{i}$
as the one component case.

Consider  the action
\begin{equation}
S=\int dt\,dx{\cal L}=\int dt\,dx
\left [ \frac{i\hbar}{2}\sum_{i=1}^{N}(\phi^{*}_{i}\partial_{t}{\phi_{i}}-
{\phi_{i}}\partial_{t}{\phi_{i}}^{*})-
\frac{\hbar^2}{2m}\sum_{i=1}^{N}\left|\left(\partial_x +i{\lambda\over 2}\rho^2
\right) \phi_{i}\right|^{2}-V\right ],
\label{eq:14}
\end{equation}
where $\rho^2=\sum_{k=1}^{N}\hat\phi_{k}\hat\phi^*_{k}$.
The equation of motion is
\begin{equation}
\ii\hbar \partial_t \phi_{i} = - {\hbar^2 \over 2m} 
\left(\partial_x +\ii\frac{\lambda}{2} \rho^2
\right )^2 \phi_{i}+\frac{\lambda\hbar}{2} (\sum_{k=1}^{N}J_{k}) \phi_{i}+V'\phi_{i},
\label{eq:11}
\end{equation}
The relation between (\ref{eq:8}) and (\ref{eq:11}) 
is 
the gauge equivalent  

\begin{equation}
\phi_{i}=
\exp\left[ \ii\frac{\lambda}{2} \int^x_{x^0}dy(\sum_{k=1}^{N}
\hat\phi^*_{k}\hat\phi_{k})(y,t)\right]\hat\phi_{i}.
\nonumber 
\end{equation}

The current corresponds to 
\begin{equation}
J_{k}={\hbar \over 2\ii m} 
\left( \phi^{*}_{k} 
\left(\partial_x +\ii \frac{\lambda}{2}\rho^2\right) \phi_{k} -
\phi_{k} \left(\partial_{x} - \ii \frac{\lambda}{2} \rho^2\right) \phi^{*}_{k}
\right).
\end{equation}

\setzero
\section{Constants of Motion and Symmetries}

We look for the conservation laws:
\begin{equation}
\frac{\partial D_{j}}{\partial t}+\frac{\partial F_{j}}{\partial x}=0,
\end{equation}
where $D_{j}$ and $F_{j}$ are respectively conserved density and flux.
There is the conserved quantities like amplitude,
\begin{equation}
D_{1}^{ij}=\phi^{*}_{i}\phi_{j},
\;\;\;
F^{ij}_{1}={\hbar \over 2im} 
\left( \phi^{*}_{i} D_x
\phi_{j} -
\hat\phi_{j}  (D_x
\phi_{i})^{*}
\right),
\label{cd}
\end{equation}
where $D_x$ stands for the ``covariant'' derivative $\partial_x
+\ii \lambda\rho^2/2 $.
In the case $i=j$ conserved density and the flux 
become the amplitude and the current respectively.
The conserved density in each line and inter-component
are only $D_{1}^{ij}$.
The other conserved densities are   sum over all components.
It is also true in the integrable model 
as (\ref{cnls}) which has infinite number of
conserved quantities.

Space translation invariance ensures momentum conservation, and 
the momentum density reads
\begin{equation}
D_{2}={\cal P}=m \sum_{k=1}^{N}J_{k}-\hbar {\lambda\over 2} \rho^4.
\label{eq:13b}
\end{equation}
With the momentum density $D_{2}$  we obtain the momentum flux 
\begin{equation}
F_{2}
= {\hbar^2\over m}\sum_{k=1}^{N}\left| D_x  \phi_{k} \right|^2
-{\hbar^2\over 4m}\partial^{2}_{x}\rho^2
+V^{'}\rho^2.
\end{equation}
Here we define ${\cal P}_{i}$ 
\begin{equation}
{\cal P}_{i}=
m J_{i}-\hbar {\lambda\over 2} \rho^2 \phi_{i}\phi_{i}^{*}.
\end{equation}
Note  that $ {\cal P}_{i}$
is not conserved quantity and the sum of $ {\cal P}_{i}$
 over all components  is the momentum density.

Energy is conserved as a consequence of time transition invariance. 
Evidently the Hamiltonian density is
\begin{equation}
D_{3}={\cal H}=\Biggl[\frac{\hbar^2}{2m}\sum_{k}^{N}\left|
D_x\phi_{k} \right|^2 +V\Biggr].
\label{eq:13a}
\end{equation}
With the energy  density $D_{3}$  there is associated  the energy  flux
\begin{equation}
F_{3}=
-\frac{\hbar^2}{2m}\sum_{k=1}^{N}
\left [ D_x \phi_{k}
\partial_t\phi^*_{k}
+(D_x \phi_{k})^* \partial_t\phi_{k}
\right].
\end{equation}

Here  $D_{2}$ is not proportional to the 
current $\sum_{k=1}^{N}J_{k}$.
Then we present the usual Galileo generator
\begin{equation}
G=t\int dx{\cal P}-m\int dx x\rho^2.
\end{equation}
We find 
\begin{equation}
\label{urgo1}
\frac{d G}{d t}=\int dx ({\cal P}-m \sum_{k}^{N}J_{k})
=-\hbar{\lambda\over 2} \int dx \rho^4,
\end{equation}  
namely $G$, depending on the sign of the coupling constant, always
increases or decreases in time.
Here we introduce generators
\begin{equation}
G_{i}=t\int dx{\cal P}_{i}-m\int dx x\rho_{i}^{2}
.
\label{gali}
\end{equation}
In the same way we can obtain 
\begin{equation}
\label{urgo2}
\frac{d G_{i}}{d t}=\int dx ({\cal P}_{i}-m J_{i})
=-\hbar{\lambda\over 2} \int dx \rho^2\rho_{i}^{2},
\end{equation}
where $\rho_{i}^{2}$ is  density in $i$-th  component $\phi_{i}\phi_{i}^{*}$.
Depending on the sign of the coupling constant$, G_{i}$  always
increases or decreases in time, too.

The additional symmetry is  dilaton invariance
which can be seen in the no coupled case.
 In
fact the action (\ref{eq:14}) is unchanged under a dilation,
$t\rightarrow a^2 t,~x\rightarrow a x$, and ${\phi}(x,t)\rightarrow
a^{1\over 2}{\phi}(a^2 t, a x )$.

The generator $D$ of the scale symmetry takes the form
\begin{equation}
D=\int dx {\cal D}=t\int dx {\cal H}-{1\over 2}\int dx x {\cal P},
\label{eq:dilaton}
\end{equation}
where the density ${\cal D}=t D_{3}-{1\over 2} x D_{2}$ obeys
the continuity equation
\begin{equation}
\partial_{t}{\cal D}+\partial_{x}\left (
t~D_{3}-{1\over 2}xD_{2}
-{\hbar^2\over 8m}
\partial_{x}\rho^2\right )=0.
\label{eq:xcontD}
\end{equation}
We can remove the last term in  (\ref{eq:xcontD})
proportional to the derivative of
$\rho^2$ by adding a super potential to the  energy-momentum tensor.
In fact if we define an improved $\hat{D}$ and $\hat{F}$
\begin{equation}
\hat{D}_{3}=D_{3}-\displaystyle{{\hbar^2\over 8m}}
\partial^{2}_{x}\rho^2,\ \ \ \ \hat{F}_{3}=F_{3}-
\displaystyle{{\hbar^2\over 8m}}\partial^{2}_{x}(\sum_{k=1}^{N}J_{k}),\ \ \ \
\hat{D}_{2}=D_{2},\ \ \ \
\hat{F}_{2}=F_{2},
\label{eq:improved}
\end{equation}
we obtain
\begin{equation}
\partial_t {\cal D}+\partial_x \left (t 
\hat{D}_{3}-\frac{1}{2} x \hat{F}_{2}\right )=0.
\end{equation}
The new energy  momentum tensor  satisfies $
2\hat{D}_{3}=\hat{F}_{2}.$

We look for solutions  that possess particular symmetries 
or whose specific  functional dependence simplified 
the structure of the original equation.
In our case  a simple ansatz is to assume 
that 
the density in $i$-th component  is a function only of $x-v_{i}t$.
We expect that this choice will allow us to explore the presence of the 
soliton like solution.
In the following we shall be conserved 
with solutions that approach 
the vacuum at spatial  infinity.
This class of solutions is strongly constrained by the symmetry of the problem.
 Substituting the ansatz  into the 
continuity equation, yields
\begin{equation}
\partial_x ( - v_{i} \rho_{i}^2(x-v_{i} t)+J_{i}(x,t))=0,
\label{ce}
\end{equation}
and hence 
\begin{equation}
\label{current1}
J_{i}(x,t)= v_{i}\rho_{i}^2(x-v_{i}t) +J_i^\infty (t).
\end{equation}
From the ansatz   at spatial  infinity
we consider the case $J_i^\infty (t)=0$.
(\ref{ce}) implies that ${\cal P}_{i}$ is a function of $x-v_{i} t$.
Sum ${\cal P}_{i}$ over  all components  is the momentum density 
${\cal P}$.
The dilation charge takes the form
\begin{equation}
D=t H -\frac{1}{2}\sum_{i=1}^{N} \int^{\infty}_{-\infty} d x (x- v_{i} t) 
{\cal P}_{i}(x- v_{i} t)-
\sum_{i=1}^{N}
\frac{v_{i}}{2} t \int^{\infty}_{\infty} dx {\cal P}_{i}(x-v_{i} t)
= t (H- \sum_{i=1}^{N}\frac{v_{i}}{2} P_{i}) -\frac{1}{2}
D_0,
\end{equation}
where 
$D_0=\sum_{i}^{N}\displaystyle{\int^{\infty}_{-\infty} dx x {\cal P}_{i}(x)}$.
Here we write a conserved momentum $P=\int dx{\cal P}$ and energy
$H=\int dx {\cal H}$.

Since
$D$ is conserved and consequently time-independent we obtain
\begin{equation}
\label{HvP}
H=\sum_{i=1}^{N}\frac{v_{i}}{2} P_{i},
\end{equation}
where $P_{i}=\int dx{\cal P}_{i}$.

From the ansatz $v_{i} P_{i}/2$
is a function of $x-v_{i}t$.
Here we define 
the energy density ${\cal H}_{i}$ in the $i$-th component
\begin{equation} 
{\cal H}_{i}=\frac{\hbar^{2}}{2m}\left | D_x \phi_{i}\right|^2.
\end{equation}
Note  that $ {\cal H}_{i}$
is not conserved quantity and the sum of $ {\cal H}_{i}$
 over all components  is the energy density.
From the ansatz $H_{i}$ is a function of $x-v_{i}t$.
Then we can obtain the relation 
\begin{equation}
\label{HvP1}
H_{i}=\frac{v_{i}}{2} P_{i},
\end{equation}
where  $H_{i}=\int dx{\cal H}_{i}$


To study further properties, we introduce the ``center of mass'' coordinate
\begin{equation}
x^{CM}_{i}(t)
=\frac{\displaystyle{\int_{-\infty}^\infty dx~ x \rho^2_{i}(x,t)}}{
         \displaystyle{\int_{-\infty}^\infty dx~\rho^2_{i}(x,t)}}.
\end{equation}
This name is easily understood
if we think of $\rho^2_{i}$ as the mass density in the $i$-th component.
 Its velocity will be
\begin{equation}
v^{CM}_{i}=\dot x^{CM}_{i}(t)
=\frac{\displaystyle{\int_{-\infty}^\infty dx~ J_{i}(x,t)}}{N_{i}},
\ \ \ \ \ {\rm with}\ \ N_{i}\equiv\int^{\infty}_{-\infty} dx ~\rho^2_{i}(x,t).
\end{equation}

Here we have used the continuity equation for the current to eliminate the
time derivative of the density.

From  (\ref{urgo1}) we can obtain  a suggestive form,
\begin{equation}
\label{purgo1}
\lambda (P_{i}- m N_{i} v^{CM}_{i}(t))
=-\hbar\frac{\lambda^2}{2}\int^{\infty}_{-\infty}
dx ~\rho^2(x,t)\rho_{i}^{2}\le 0.
\end{equation}
Being valid for all $t$, this implies
\begin{equation}
\lambda P_{i}\le m N_{i} \lambda \min_{t\in  {\rm R}}\{ v^{CM}_{i}(t)\}.
\end{equation}
  One can show an analogous inequality for the energy. In
fact let us consider the following inequality
\begin{equation}
\int^\infty_{-\infty} dx 
\left | \phi_{i} + w \frac{\hbar}{2 m \ii}D_x \phi_{i}\right|^2
\ge 0,
\end{equation}
where $w$ is  an arbitrary parameter. In terms of the physical quantities
\begin{equation}
N_{i} + w N_{i} v^{CM} (t)+w^2 \frac{E_{i}}{2m} \ge 0,
\end{equation}
with $E_{i}=\displaystyle{\int^\infty_{-\infty} dx H_{i}}$.
 The fact that the
previous equation holds for all $w$  entails
\begin{equation}
E_{i}\ge \frac{m N_{i} (v^{CM}_{i})^{2}}{2}.
\end{equation}
Note that here we consider the case the potential is absent.
 From the ansatz we can obtain  that $v^{CM}_{i}(t)=v_{i}$. 
Thus     (\ref{purgo1})
can be written as
\begin{equation}
\lambda v_{i}
\left( H_{i}-\frac{m N_{i} v^2_{i}}{2}\right)\le 0,
\end{equation}
where we used  (\ref{HvP1}).

It implies $\lambda v_{i}<0$,
{\it i.e.} the soliton is ``chiral''. 
If we set $\lambda>0$, it shows that all the traveling waves 
move to the left.

\setzero
\section{Explicit solution}

The coupled nonlinear Schr\"{o}dinger (coupled NLS) equation \cite{Ma}
is
\begin{equation}
\ii\partial_{T}\hat\phi_{i}+\partial_{X}^2\hat\phi_{i}
+2
(\sum_{k=1}^{N}\kappa_{k}\rho_{k})\hat\phi_{i}=0,
\label{cnls}
\end{equation}
where use the normalized time and space variables
$t=\hbar T$ and $x=\sqrt{\hbar^{2}/2m}X$.
This equation has bright solutions only in  all 
$\kappa_{i}>0$.
If the $\kappa_{i}< 0$ 
then the soliton solution in the $i$-th component becomes dark.\cite{hh} 
It possesses the famous soliton solution of the form
\begin{equation}
\phi_{i}=\frac{g_{i}}{f},\;\;\;\phi_{i}^{*}=\frac{g_{\bar{i}}}{f},
\label{fg}
\end{equation}
for $i=1,2,\cdots,N$.
$f$ is a real function.
Here we consider the simple case $N=2$.
The solution which has  one soliton 
for each component is expressed by
\begin{eqnarray}
g_{1}&=&\exp[\eta_{1}]+\alpha_{12}\alpha_{2\bar{2}}\alpha_{\bar{2}1}
\beta_{2\bar{2}}\exp[\eta_{1}+\eta_{2}+\eta_{\bar{2}}],
\nonumber \\
g_{\bar{1}}&=&\exp[\eta_{\bar{1}}]+\alpha_{\bar{1}2}\alpha_{2\bar{2}}\alpha_{\bar{2}\bar{1}}
\beta_{2\bar{2}}\exp[\eta_{\bar{1}}+\eta_{2}+\eta_{\bar{2}}],
\nonumber \\
g_{2}&=&\exp[\eta_{2}]+\alpha_{21}\alpha_{1\bar{1}}\alpha_{\bar{1}2}
\beta_{1\bar{1}}\exp[\eta_{2}+\eta_{1}+\eta_{\bar{1}}],
\nonumber \\
g_{\bar{2}}&=&\exp[\eta_{\bar{2}}]+\alpha_{\bar{2}1}
\alpha_{1\bar{1}}\alpha_{\bar{1}\bar{2}}
\beta_{1\bar{1}}\exp[\eta_{\bar{2}}+\eta_{1}+\eta_{\bar{1}}],
\nonumber \\
f&=&1+\alpha_{1\bar{1}}\beta_{1\bar{1}}\exp[\eta_{1}+\eta_{\bar{1}}]
+\alpha_{2\bar{2}}\beta_{2\bar{2}}\exp[\eta_{2}+\eta_{\bar{2}}]
\nonumber \\
& &
+\alpha_{1\bar{1}}\alpha_{12}\alpha_{\bar{2}1}
\alpha_{2\bar{2}}\alpha_{\bar{1}2}\alpha_{\bar{2}\bar{1}}
\beta_{1\bar{1}}\beta_{2\bar{2}}
\exp[\eta_{1}+\eta_{\bar{1}}+\eta_{2}+\eta_{\bar{2}}],
\label{solution}
\end{eqnarray}
with
\begin{eqnarray}
\eta_{\mu}&=&p_{\mu}X+\ii p_{\mu}^{2}T+\eta_{\mu}^{0},
\nonumber \\
\alpha_{\mu\nu}&=&\frac{p_{\mu}-p_{\nu}}{p_{\mu}+p_{\nu}},
\nonumber \\
\beta_{\mu\bar{\mu}}&=&\frac{\kappa_{\mu}}
{p_{\mu}^{2}-p_{\bar{\mu}}^{2}}.
\end{eqnarray}
$p_{\mu}$ and $\eta_{\mu}$ are complex constant 
parameters related to the amplitudes and position of solitons
 respectively for $\mu=1,\bar{1},2, \bar{2}$.
$p_{\mu}$,  $p_{\bar{\mu}}$ 
and 
$\eta^{0}_{\mu}$, $\eta^{0}_{\bar{\mu}}$
are the complex conjugate.
Note  that $\kappa_{\mu}=\kappa_{\bar{\mu}}$.
Here after we choose the parameter 
\[
\eta^{0}_{\mu}=\log \frac{\kappa_{\mu}}{4a_{\mu}^{2}}.
\]
If we set $p_{\mu}=a_{\mu}+{\rm i}b_{\mu}$, 
the speed of soliton $V_{\mu}$ is 
presented as
\begin{equation}
\frac{V_{\mu}}{2}=b_{\mu}.
\label{speed}
\end{equation}
To find the solution of (\ref{eq:11}) we take the form (\ref{fg})
so that
\begin{equation}
j_{i}=\frac{\hbar}{2{\rm i}m}
\frac{g_{i}'\bar{g}_{i}-\bar{g}_{i}'g_{i}}{f^{2}}
=
\frac{\hbar b_{i}g_{i}\bar{g}_{i}}{mf^{2}}.
\label{relation}
\end{equation}
where $'$ means differentiation with space 
and our equation (\ref{eq:11}) becomes identical 
(\ref{cnls}) 
with
\begin{equation}
\kappa_{i}=-\frac {\hbar^{2}\lambda V_{i}}{2}.
\label{c}
\end{equation}
The last equation can be obtained using (\ref{speed}).
So that (\ref{solution}) with the condition (\ref{c})
becomes the soliton solution of  (\ref{eq:11}).
Note that $V_{i}$ is a speed in the coordinate $X$ and $T$.
The soliton solution exists provided $\kappa_{i}>0$;
 this requires 
\begin{equation}
V_{i}<0\;\;\;\;({\rm for}\;\;{\rm  all }\;\;\;i\;\;\;{\rm if   }\;\;\lambda>0).
\end{equation}
The soliton in all components  can only move to the left.
Had we taken $\lambda<0$, then the solitons  could move
only to the right. 

We now discuss the phase shifts.
We assume $V_{1}> V_{2}>0$.
In the limit $t\rightarrow -\infty$
with $\eta_{1}$ and $\eta_{\bar{1}}$ fixed
and $\eta_{2}$ and $\eta_{\bar{2}}\rightarrow -\infty$,
the soliton solution approaches to
\begin{equation}
\phi_{1}=\frac{\exp[\eta_{1}]}{1+\alpha_{1\bar{1}}\beta_{1\bar{1}}
\exp[\eta_{1}+\eta_{\bar{1}}]}
=a_{1}\sech a_{1}(X-V_{1}T)
\exp[\ii\frac{V_{1}}{2}X-\ii(\frac{V_{1}^{2}}{4}-a_{1})T],
\end{equation}
which is the 1-soliton  solution moving with the velocity $V_{1}$
 in the one component case.
Also the other   limit $t\rightarrow \infty$
with $\eta_{1}$ and $\eta_{\bar{1}}$ fixed
and $\eta_{2}$ and $\eta_{\bar{2}}\rightarrow \infty$,
the  result is 
\begin{eqnarray}
\phi_{1}&=&\frac{\alpha_{12}\alpha_{\bar{2}1}
\exp[\alpha_{1\bar{1}}\beta_{1\bar{1}}\eta_{1}]}
{1+\alpha_{1\bar{1}}\alpha_{12}\alpha_{\bar{2}1}
\alpha_{2\bar{2}}\alpha_{\bar{1}2}\alpha_{\bar{2}\bar{1}}
\beta_{1\bar{1}}\beta_{2\bar{2}}
\exp[\eta_{1}+\eta_{\bar{1}}]}
\nonumber \\
&=&
a_{1}\sech a_{1}(X-V_{1}T+{\rm Re}\Delta_{1})
\exp[\ii\frac{V_{1}}{2}X-\ii(\frac{V_{1}^{2}}{4}-a_{1})T
+\ii{\rm Im}\Delta_{1}],
\end{eqnarray}
where 
$\Delta_{1}$ is the phase shift which is the effort of the 
inter-component interactions.
One can  repeat the same limiting procedure but with 
$\eta_{2}$ and $\eta_{\bar{2}}$ fixed and obtain the other 
soliton sector moving with velocity $V_{2}$.
This shows that 
the solution describes the  scattering of the chiral solitons.

From the solutions 
we can obtain the phase shifts
\begin{eqnarray}
\Delta_{1}&=&\log\alpha_{12}\alpha_{\bar{2}1},
\;\;\;
\Delta_{\bar{1}}=\log\alpha_{\bar{1}2}\alpha_{\bar{2}\bar{1}},
\nonumber  \\
\Delta_{2}&=&\log\alpha_{\bar{1}2}\alpha_{21},
\;\;\;
\Delta_{\bar{2}}=\log\alpha_{\bar{1}\bar{2}}\alpha_{\bar{2}1},
\end{eqnarray}
where $\Delta_{i}$ is the phase shift of the $i$-th component.

Here we change the ``charge density'' $\rho^2$
\begin{equation}
\rho^{2}=\rho_{1}^2-\rho_{2}^2.
\end{equation}
The equation of motion  becomes 
\begin{equation}
i\hbar\partial_{t}\hat\phi_{i}+{\hbar^2\over 2m}\partial_{x}^2\hat\phi_{i}-\hbar 
{\lambda}(j_{1}-j_{2})\hat\phi_{i}=0.
\label{cv} 
\end{equation}
To find the solution of (\ref{cv})
 we take also the form (\ref{fg}).
Using  (\ref{relation}) 
we can obtain the relations
\begin{equation}
\kappa_{1}=-\frac {\hbar^{2}\lambda V_{1}}{2},\;\;\;\;
\kappa_{2}=\frac {\hbar^{2}\lambda V_{2}}{2}.
\end{equation} 
We can obtain the explicit soliton solution 
which  can move left in the component 1  and 
right in the component 2, in the case $\lambda>0$. 

\setzero
\section{Integrable Coupled Derivative Nonlinear 
Scr\"{o}dinger equation}
\newcommand{\eqref}[1]{eq.(\ref{#1})}
\newcommand{\secref}[1]{\ref{#1}}
\newcommand{\beq}{\begin{equation}}
\newcommand{\eeq}{\end{equation}}
\newcommand{\bl}[1]{\makebox[#1em]{}}
\newcommand{\pa}{\partial}
\newcommand{\ovl}[1]{\overline{#1}}
\newcommand{\ul}[1]{\underline{#1}}

\newcommand{\qtil}{\hat{q}}
\newcommand{\Qtil}{\hat{Q}}
\newcommand{\ftil}{\tilde{f}}
\newcommand{\gtil}{\tilde{g}}
\newcommand{\htil}{\tilde{h}}
\newcommand{\uhat}{\hat{u}}
\newcommand{\vhat}{\hat{v}}

\newcommand{\ee}{\mbox{e}}
It is well known that
the coupled hybrid nonlinear Schr\"{o}dinger equation\cite{h1}\cite{h2}
\beq
\ii \Qtil_T^{i} + \Qtil_{XX}^{i} + \sum_{k}^{N}\beta|\Qtil ^{k}|^2 \Qtil^{i}
 + \ii\alpha\left(  \sum_{k}^{N}|\Qtil|^{2k}\Qtil ^{k}\right)_X = 0,
\label{hc}
\eeq
 is integrable.
These equations are a hybrid of the coupled NLS \cite{Ma}
and coupled derivative NLS  equation.\cite{h3}
If we set
\beq
Q^{i}=\Qtil ^{i}\exp\left( -2\ii\delta\sum_{k}^{N}
\int^{X}|\Qtil^{k}|^2 \mbox{d}X \right),
\label{eqn:GT}
\eeq
then (\ref{hc}) is gauge-equivalent \cite{WS}\cite{h3} to
\begin{equation}
\ii Q _T ^{i}+ Q _{XX}^{i} + \beta \rho^{2}_{Q} Q^{i}
-2\ii\delta AQ^{i}+\ii(2\delta+\alpha)BQ^{i}+\ii(4\delta+\alpha)
\rho^{2}_{Q}Q^{i}_{X}
 + \delta(4\delta+\alpha)\rho^{4}_{Q}  Q^{i} =0,
\label{eqn:GDNLS0}
\end{equation}
where
\begin{eqnarray}
A&=&\sum_{k}^{N}(Q_{X}^{k}Q^{k*}-Q^{k*}_{X}Q^{k}),
\nonumber \\
B&=&\sum_{k}^{N}(Q_{X}^{k}Q^{k*}+Q^{k*}_{X}Q^{k}),
\nonumber \\
\rho^{2}_{Q}&=&\sum_{k}^{N}|Q^{k}|^2,
\end{eqnarray}
(\ref{eqn:GDNLS0}) is the coupled version of the generalized 
derivative NLS  equation.\cite{Ku}\cite{Ka}

If we set $4\delta+\alpha=0$, we can obtain 
the coupled version  of Chen-Lee-Liu type equation\cite{DNLS2}
\begin{equation}
\ii Q_T^{i} + Q_{XX}^{i}+  \beta \rho^{2}_{Q} Q^{i}
-4\ii\delta(\sum_{k}^{N}(Q_{X}^{k}Q^{k*})Q^{i}
 = 0.
\label{eqn:CCL}
\end{equation}


In one component case 
there is the relation
\begin{equation}
-AQ=BQ-2\rho_{Q}^{2}Q_{X},
\label{ur}
\end{equation}
so that 
 by adding an attractive potential 
with a fixed coefficient (in the case $3\delta+8\alpha=0$)
the chiral soliton model can be made integrable.\cite{COREA}.
On the other hand 
in multi-component case 
we can not use (\ref{ur}),
then 
only adding potential
we can not make the model integrable.

\setzero
\section{Concluding Remarks}
We have studied a family of 1+1 dimensional 
theories  
that describes non relativistic  bosons interacting with 
a gauge potential in a multi-component  model.
This form was suggested from the dimensional 
reduction of Chern-Simons theory coupled  to
non-relativistic 
matter and  it represents a simple way to introduce chiral excitations
as the one component case.
It can be exactly reduced, solving for $A_{\mu}$ and $B$, to a bosonic 
theory with self and inter-component interactions.
For this model a local Lagrangian formulation is possible.
The soliton structure of the theory has been examined, and it 
exhibits an interesting chiral behavior.
In the multi-component case 
we can see the soliton-soliton scattering 
for the inter-component interactions.
If we change the density, 
we can choose the chirality of the 
soliton solution in each component.
 Only adding potential 
we can not make the model integrable.

\end{document}